\documentclass[prc,preprint,superscriptaddress,showpacs,amssymb,amsmath,amsfonts,aps]{revtex4}

\usepackage{graphicx}
\usepackage{dcolumn}
\usepackage{bm}
\usepackage{epsfig}

\newcommand{\jpsi}{$J/\psi$\ }

\begin{document}


\title{Search for Sub-threshold Photoproduction of \jpsi Mesons}

\newcommand*{\JLAB}{Thomas Jefferson National 
Accelerator Facility, Newport News, Virginia 23606}
\newcommand*{\MSS}{Mississippi State University, Mississippi 
State, Mississippi 39762 }
\newcommand*{\WITS}{University of the Witwatersrand, 
Johannesburg, South Africa }
\newcommand*{\ARGONNE}{Physics Division, Argonne National 
Laboratory, Argonne, Illinois 60439 }
\newcommand*{\YERPHY}{Yerevan Physics Institute, Yerevan, Armenia}
\newcommand*{\MARYLAND}{University of Maryland, College Park, 
Maryland 20742 }
\newcommand*{\FIU}{Florida International University, 
University Park, Florida 33199 }
\newcommand*{\HAMPTON}{Hampton University, Hampton, Virginia 23668 }
\newcommand*{\MIT}{Massachusetts Institute of Technology, Cambridge, Massachusetts, 02139}
\newcommand*{\HOU}{University of Houston, Houston, TX 77204 }
\newcommand*{\UVA}{University of Virginia, Charlottesville, 
Virginia 22904 }
\newcommand*{\BASEL}{University of Basel, Basel, Switzerland}
\newcommand*{\PSI}{Paul Scherrer Institut, Switzerland}
\newcommand*{\NORFOLK}{Norfolk State University, Norfolk, Virginia, 23504}
\newcommand*{\UJO}{University of Johannesburg, 
Johannesburg, South Africa }
\newcommand*{\UNH}{University of New Hampshire, Durham, New Hampshire 03824}
\newcommand*{\USJ}{Universidad Metropolitana, San Juan, PR 00928}
\newcommand*{\PENN}{University of Pennsylvania, University Park, PA 16802}

\author{P.~Bosted} 
\affiliation{\JLAB}
\author{J.~Dunne} 
\affiliation{\MSS}
\author{C.~A.~Lee} 
\affiliation{\WITS}
\author{P.~Junnarkar} 
\affiliation{\MSS}
\author{M.~Strikman} 
\affiliation{\PENN}
\author{J.~Arrington} 
\affiliation{\ARGONNE}
\author{R.~Asaturyan} 
\affiliation{\YERPHY}
\author{F.~Benmokhtar} 
\affiliation{\MARYLAND}
\author{M.E.~Christy} 
\affiliation{\HAMPTON}
\author{E.~Chudakov} 
\affiliation{\JLAB}
\author{B.~Clasie} 
\affiliation{\MIT}
\author{S.~H.~Connell} 
\affiliation{\UJO}
\author{M.~M.~Dalton} 
\affiliation{\WITS}
\author{A.~Daniel} 
\affiliation{\HOU}
\author{D.~Day} 
\affiliation{\UVA}
\author{D.~Dutta} 
\affiliation{\MSS}
\author{R.~Ent} 
\affiliation{\JLAB}
\author{N.~Fomin} 
\affiliation{\UVA}
\author{D.~Gaskell} 
\affiliation{\JLAB}
\author{T.~Horn} 
\affiliation{\MARYLAND}
\affiliation{\JLAB}
\author{N.~Kalantarians}
\affiliation{\HOU}
\author{C.E.~Keppel} 
\affiliation{\HAMPTON}
\author{D.G.~Meekins} 
\affiliation{\JLAB}
\author{H.~Mkrtchyan} 
\affiliation{\YERPHY}
\author{T.~Navasardyan} 
\affiliation{\YERPHY}
\author{J.~Roche} 
\affiliation{\JLAB}
\author{V.~M.~Rodriguez} 
\affiliation{\HOU}
\affiliation{\USJ}
\author{D.~Kiselev~(nee~Rohe)} 
\affiliation{\BASEL}
\affiliation{\PSI}
\author{J.~Seely} 
\affiliation{\MIT}
\author{K.~Slifer} 
\affiliation{\UVA}
\affiliation{\UNH}
\author{S.~Tajima} 
\affiliation{\UVA}
\author{G.~Testa} 
\affiliation{\BASEL}
\author{Roman Trojer} 
\affiliation{\BASEL}
\author{F.R.~Wesselmann} 
\affiliation{\NORFOLK}
\author{S.A.~Wood}  
\affiliation{\JLAB}
\author{X.C.~Zheng} 
\affiliation{\UVA}
\date{\today}

\begin{abstract}
A search was made for sub-threshold $J/\psi$ production 
from a carbon target using
a mixed real and quasi-real Bremsstrahlung photon beam with
an endpoint energy of 5.76 GeV. No events were observed, which 
is consistent with predictions assuming quasi-free
production. The results place limits on exotic mechanisms that
strongly enhance quasi-free production.  
\end{abstract}

\maketitle

\section{Introduction}
\indent One of the main goals of nuclear physics is to understand to what 
extent a nucleus differs from a loosely bound system of 
quasi-independent nucleons. When nucleons are very close spatially,
corresponding to rare high momentum components of the single
particle wave function, many interesting and potentially exotic
configurations can arise. One way to look for such configurations
is with reactions that are significantly sub-threshold to
production from a free nucleon. Of all such reactions, photoproduction
of charmonium is one of the cleanest because the charm quark
content of a nucleon is expected to be small 
compared to the light quarks. 
In light meson photoproduction, the quark content of the mesons can
originate in the nuclear target, while in the case of charmonium
photoproduction, the quark-interchange mechanism is essentially
absent, and the reaction must proceed via gluon exchange in order
for color to be conserved. In addition, the heavy mass of the
charm quark (about 1.5 GeV) ensures a hard scale to the problem,
making it more tractable in QCD. 

\subsection{Kinematics of sub-threshold photoproduction}
\begin{figure}[hbt]
\includegraphics[width=4in]{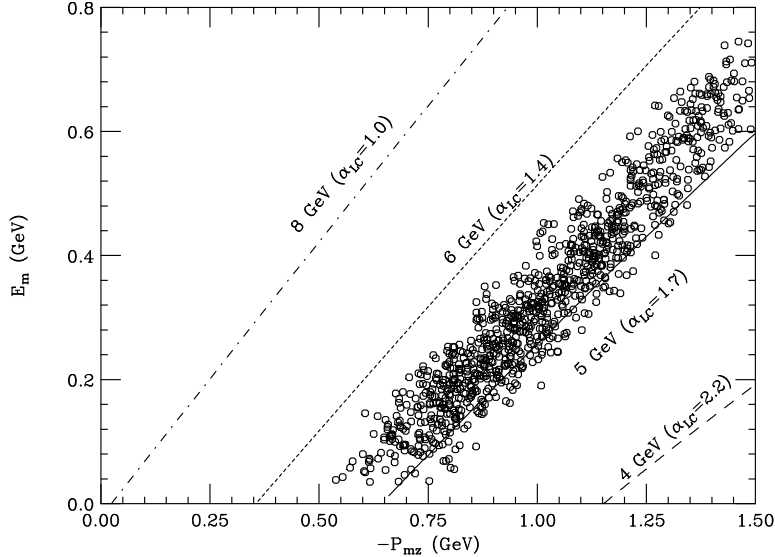}
\caption{The four curves show the maximum value of missing
energy $E_m$ as a function of the negative $z$ component of missing
momentum $P_m$, for 4, 5, 6, and 8 GeV photons, for the 
case of quasi-free photoproduction of \jpsi mesons from 
a nucleon bound in a nucleus. The circles
represent the relative distribution of events predicted for the 
present experiment for the specific model described in the
text [although no events were actually observed]. The values of
light cone fraction $\alpha_{LC}$ are indicated for each photon energy.}
\label{fig:pmem}
\end{figure}

The goal of this experiment was to study the production mechanisms 
in the extreme conditions of matter
that may be relevant in heavy ion collisions. 
These  conditions are ensured by using a photon beam 
energy well below photoproduction threshold on a free nucleon. 
In the quasi-free picture where a photon 
interacts with a single off-shell
nucleon in a nucleus, the nucleon three-momentum 
$\vec P_m$ must be pointing
anti-parallel to the photon direction ($z$) for the invariant mass
of the photon-nucleon system $s$ to be above the threshold value
of $(m + M_j)^2=16.3$ GeV$^2$, where $m$ is the nucleon mass
and $M_j$ is the \jpsi mass. An additional constraint is that the
missing energy $E_m$ cannot be too high.
Specifically, we calculated $s$ for a given photon energy $k$ using
\begin{equation}
\label{eq:s}
s = (k + m - E_m)^2 - k^2 - P_m^2  - 
2  \vec k \cdot  \vec P_m.
\end{equation}
From Eq.~\ref{eq:s}, it is clear that larger values of $E_m$
correspond so smaller values of $s$, for fixed values of $P_m$.
The highest value of $E_m$ that is kinematically allowed 
is shown as a function of the magnitude of $\vec P_m$ for photon
energies of 4, 5, 6, and 8 GeV in Fig.~\ref{fig:pmem} (for the case
of $\vec P_m$ anti-parallel to the photon direction).
Since $E_m$
must be greater than zero, this leads to the conclusion that
for \jpsi photoproduction, the minimum nucleon momentum $P_m$ increases
 as the photon energy decreases. In particular the minimum
momentum for 4 GeV photons is 1.15 GeV, for 6 GeV photons it is
0.35 GeV, and for 8 GeV photons it is 0.05 GeV.
As discussed below, it is generally thought that the region
where few-nucleon short-range  correlations, hidden color configurations,
and other short-range  effects play a significant role 
for momenta larger than  0.35 GeV~\cite{laget}, 
corresponding to a photon 
energy of 6 GeV. This was therefore chosen as the ideal
energy for the present experiment. In practice, we used a
slightly lower energy due to accelerator limitations at the
time that the data were taken.

Another way to realize the importance of short-range 
correlations comes from the observation that the photon
threshold to produce \jpsi meson is 5.7 GeV for a deuteron
target at rest, and 4.8 GeV for a triton or $^3$He target. Thus,
two-nucleon and particularly three-nucleon correlations 
in the carbon nucleus are needed to kinematically permit
\jpsi photoproduction with photons with energies below 
6 GeV. In the language of light-front (infinite momentum frame) 
QCD, photon energies
of 4, 6, and 8 GeV correspond to minimal light-cone fractions
$\alpha_{LC}=2.2$,~1.4, and 1.0 respectively, where 
$\alpha_{LC}= 1 - (E_m - P_{mz})/m$. In the limit of very large 
quark masses the cross section of onium production in the 
impulse approximation is expressed through the light-cone 
density matrix of the nucleus as $s$ in Eq.~\ref{eq:s} is 
$\approx k\alpha_{LC} m$. 
For the \jpsi case constraints due to the recoil mass contribute 
as well. 

The kinematics of the
present experiment (represented by the open circles in 
 Fig.~\ref{fig:pmem}) all lie in the range $1.5<\alpha_{LC}<2$,
corresponding to the region of multi-nucleon 
correlations~\cite{Strikman}.
The recoil effects in these kinematics  essentially 
remove the contribution of
two-nucleon short range correlations, as in this case the recoil 
energy is taken by one nucleon (this is the kinematic domain which 
was studied in recent Jlab experiments using the  $(e,e'NN)$ 
reaction~\cite{Eli1} 
and inclusive scattering~\cite{Kim1} at $1< x < 2$). 
The dominant contribution originates from the region where at 
least two nucleons balance the struck nucleon momentum, i.e. 
three or more nucleon correlations, which were studied 
so far only in $x >2$ measurements~\cite{Kim2}.

\subsection{The high $\vec P_m$ region}
The high $\vec P_m$  momentum 
region is where few-nucleon short-range correlations are known to
be important~\cite{Hipm,frst}, potentially leading to 
significantly enhanced  yields compared to a simple
quasi-free model. 
In the hard-scattering picture~\cite{laget}, this
could correspond to strong contributions of
twist 3-gluon exchange, compared to the minimal 2-gluon exchange
needed to ensure the \jpsi color singlet final state.
The influence of intrinsic charm contributions, or hidden color
contributions, could potentially increase the cross section by
an order of magnitude above the expectations of a 
quasi-free model (see Fig. 5 of Ref.~\cite{laget}). 
Another possibility to enhance the sub-threshold cross section is
a diagram in which two gluons are exchanged to two different
nucleons. This type of process could become important in the
sub-threshold region, because each of the gluons could have a much
lower momentum fraction than if the pair of gluons came from
a single nucleon. 

\subsection{Relation to sub-threshold hadroproduction}
In $pA$ collisions, it has been observed that
anti-protons and kaons  are produced on nuclear
targets at substantially lower energies than is kinematically possible
on free nucleons~\cite{Ca89}. Scattering on a single nucleon in the
nucleus would, at these energies, require a single-nucleon
momentum in the vicinity of 800 MeV. While the $pA$ data 
can be fit assuming such high momenta are sufficiently likely, 
this assumption leads to an underestimate of
sub-threshold production in $AA$ collisions by about three orders of
magnitude~\cite{Shor}.

There are at least two qualitatively different scenarios for the
observed sub-threshold production of anti-protons~\cite{Hoyer}. 
One scenario is that 
the projectile strikes a local ``hot spot" with
a high energy density in the nucleus. The effective mass of the
scatterer is high, lowering the kinematic threshold. Alternatively,
the momentum required to create the anti-proton
is not transferred locally, but picked up in an extended longitudinal
region. Establishing either scenario would teach us
something qualitatively new about rare, highly excited modes of the
nucleus.

Sub-threshold photoproduction can help to 
identify the  correct mechanism, because 
the $c\overline{c}$ component of the photon is almost point-like
at the charm threshold and below. Effects due to the
shrinking effective size of a hadron probe near threshold are
eliminated. The $c\overline{c}$ pair is created locally, within a proper
time $\tau \simeq 1/m_c$. The extended acceleration scenario
is thus not effective for charm
photoproduction. If significant sub-threshold charm photoproduction
occurs (beyond what can be ascribed to the quasi-free model) this
selects the hot spot scenario.

\section{Quasi-free modeling}
To model quasi-free photoproduction from a heavy nucleus, 
we use a convolution integral similar to the model for the deuteron 
described in Ref.~ \cite{laget}
\begin{equation}
\label{eq:dsig}
d\sigma=\int \Phi(k) dk \int 
\frac{d\sigma_0(s,t)}{dt}\, \alpha_{LC}(P_m,E_m) \,
S(E_m,\vec P_m)d^3 \vec P_m dE_m dt
\end{equation}
where the integral over nucleon momentum $\vec P_m$  and 
energy $E_m$ is limited to the kinematically allowed region, 
as described above, 
and the photon-nucleon center of mass energy squared $s$ and 
momentum transfered $t$ are functions of $E_m$,  $\vec P_m$,
and photon energy $k$. The elementary amplitude is 
far off shell ($p^2$ for the interacting nucleon 
is $\ll m^2$), which may lead of a significant suppression 
of the cross section, not taken into account in the present analysis.
The photon flux $\Phi(k)$ is given by 
$a/k$, where $a$ in the present experiment is the sum
of the target thickness in radiation lengths divided by two,
and an effective quasi-real electroproduction factor~\cite{Dalitz} of 0.02. 
The cross section $d\sigma_0(s,t)/dt$ is for \jpsi photoproduction
from a free nucleon, and we 
assume for simplicity that neutron and proton cross sections
are equal. The flux term~\cite{Strikman} 
\begin{equation}
\label{eq:flux}
\alpha_{LC}(P_m,E_m)=\big(1 - \frac{E_m}{m} - \frac{\vec k 
\cdot \vec P_m}{km}
\big)
\end{equation}
averages to about 1.7 for the kinematics of this experiment (see
Fig.~\ref{fig:pmem}).
The function $S(E_m,\vec P_m)$ is 
a carbon spectral function~\cite{Benhar}.  The spectral 
function is defined as the probability of finding a 
nucleon of momentum $\vec{P}_m$ and separation energy 
$E_m$ in the nucleus. In this picture, the only unknowns
are the model for $d\sigma_0(s,t)/dt$ and the 
carbon spectral function at high $\vec P_m$
and $E_m$. 


\section{Existing data and models near threshold}
There are no published data on sub-threshold photoproduction, but
it is useful to review the existing data just above threshold on
a free nucleon, as this provides the baseline for sub-threshold
predictions. The existing data below 20 GeV 
come from Cornell \cite{Cornell}
using 9.3 to 11.8 GeV photons, and from SLAC \cite{SLACDA} from
13 to 21 GeV. These experiments detected lepton pairs from the
\jpsi decay to provide relatively background-free measurements.
Additional unpublished data from SLAC \cite{SLACSA} detected only
a single lepton, leading to relatively large background subtractions.

The experiments typically parameterize the data according to
$d\sigma/dt=Ae^{bt}$. 
What is remarkable is that, while $b$ has values
of 3 to  5 GeV$^{-2}$ at high energy, which are characteristic
of diffractive processes, the values drop rapidly near threshold,
with Cornell \cite{Cornell} quoting a value of 
only $1.25\pm0.2$  GeV$^{-2}$ for an
11 GeV photon energy. It was pointed out by Ref.~\cite{frankfurt}, the
actual slope of the data seems to be more like 1.5 GeV$^{-2}$, still
quite small. The Cornell
value is more than a factor of two below the SLAC value of 
$b=2.9\pm0.3$  GeV$^{-2}$ at 19 GeV. 
It is difficult to reconcile the two experiments with a 
smooth fit, assuming the exponential form corresponds to an
effective form factor. One way to resolve this
is to assume that $d\sigma/dt$ scales as a dipole form factor squared of 
the form $(1-t/m_0^2)^{-4}$ \cite{frankfurt}. A reasonably good
fit to all data up to photon energies of 100 GeV can be found 
with $m_0^2\approx 1$ GeV$^2$. Since each experiment measured over
a limited range of $t$, and $-t_{min}$ increases near threshold, 
a natural explanation for the variation of $b$ with photon energy
can be found. We found a reasonable dipole fit to the low energy
data is given by $d\sigma/dt=2.5/(1-t)^4$ nb/GeV$^2$, where $t$
is in units of GeV$^2$. In the QCD picture of Laget~\cite{laget},
this almost flat dependence on photon energy corresponds to three-gluon
exchange dominance, while 2-gluon exchange should have an
extra factor of $(1-x)^{2}$, where $x$ is the momentum fraction. 
In the definition of Ref.~\cite{laget},
$x=((m+M_J)^2-m^2)/(s-m^2)$ 
is unity for at the threshold for photoproduction. 
In the approach of Ref.~\cite{frankfurt}, a different definition of $x$
is used which yields a maximum value of about 0.8, 
so factors such as $(1-x)^2$ do not cause
a large threshold suppression. 

In summary, there is a very large uncertainty in how to extend
fits to existing free-nucleon cross section measurements below 11 GeV. 
This is discussed in more detail below in the context of cross
section limits from the present experiment.

\section{Experiment}

\begin{figure}[hbt]
\includegraphics[width=3.4in]{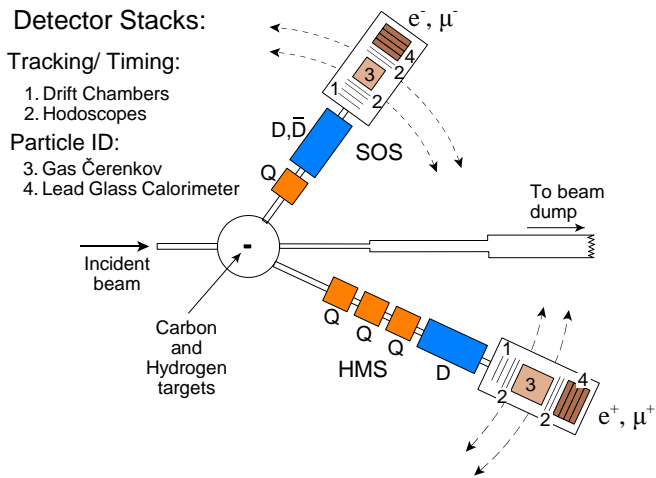}
\caption{(color online) Layout of spectrometers and detectors, where  
{\sf Q} and {\sf D} represent quadrupole and 
dipole magnets respectively.}
\label{exp_setup}
\end{figure}

The experiment was performed in 2004 in Hall C at the Thomas 
Jefferson National Accelerator facility (JLab).  The 
layout of the spectrometers and detectors is indicated 
in Fig.~\ref{exp_setup}.  

\subsection{Beam and target}
The \jpsi search data taking 
used a 5.76 GeV  continuous-wave (CW) electron beam with
a typical current of 60 $\mu$A, and lasted for eight days. 
The integrated beam charge on target was 27 Coulombs.
The electron beam impinged on a 
narrow solid $^{12}$C target with a thickness (in the beam
direction) of 2.5 gm/cm$^2$, or 0.06 radiation length (r.l.).  
The effective photon flux per electron was $0.05 dk/k$, where
$k$ is the photon energy, obtained by considering that on 
average half of the real photons produced in the target are 
usable, plus an effective 2\% from small-angle electroproduction.  

\subsection{Spectrometers and detectors}
The experiment used the High Momentum 
Spectrometer (HMS) in coincidence with the Short Orbit 
Spectrometer (SOS) to measure lepton pairs from \jpsi 
decays (6\% branching ratio for each of $e^+e^-$ and $\mu^+\mu^-$).  
The magnets of the HMS were configured to detect positive 
charged particles while the SOS had negative polarity.  
The HMS central momentum was set at 3.5 GeV/$c$ and the
central scattering angle at 24 degrees.  
The SOS had a central momentum set near its maximum value 
of 1.7 GeV/$c$ and the central angle set to 53 degrees.
These settings were chosen to optimize the acceptance for 
forward-angle \jpsi mesons decaying to lepton pairs, with
the assumption that the angular distribution of the lepton pairs
is given by $1+\cos^2(\theta_{cm})$~\cite{schilling}, where
$\theta_{cm}$ is the center-of-mass (c.m.) decay angle. 
The optimization was done assuming the elementary cross section
model had the form $d\sigma/dt=2.5/(1-t)^4$ nb/GeV$^2$ and the
spectral function of Benhar~\cite{Benhar} (with one of the
high $P_m$ extrapolations discussed below) to generate a sample
of quasi-free \jpsi mesons.
The choice of spectrometer polarities was made to minimize backgrounds
and random coincidences. The CW nature of the electron beam
was essential to reducing accidental coincidences to an
acceptable level.  


\subsection{Particle identification}
Electrons in the SOS and positrons in the HMS were identified by
requiring a signal of several photoelectrons in threshold 
gas Cherenkov counters, with the threshold momentum for pions
or heavier particles to produce Cherenkov light set above
the maximum accepted particle momentum in the given spectrometer.
In addition, at least 70\% (90\%) of the particle energy was
required to be deposited in an array of lead glass blocks
in the SOS (HMS). These cuts reduced contamination from 
other particles to a negligible level, while preserving an
efficiency of over 90\% of electron-positron pairs. 

Separating muons from pions was
more problematic. In the case of the HMS spectrometer, the pion
threshold in the gas Cherenkov counter was set just below
the maximum accepted momentum (i.e., about 8\% above the central
momentum). The threshold for muons was
25\% lower than for pions, and therefore 17\% below the
central momentum of the spectrometer (which has a nominal 
momentum acceptance of $\pm 10\%$).
With approximately  10 photo-electrons (PE) for 
fully relativistic particles (such as electrons), 
the muons produced on average 2 PE to 4 PE over the spectrometer
acceptance. With a threshold of 1 PE,
the efficiency was varied from  85\% to 98\%, depending on muon momentum. 
The original Cherenkov counter in the SOS was designed 
to be sub-atmospheric, hence could not achieve the correct 
index of refraction to trigger on muons. It was therefore
replaced with the SLAC 1.6 GeV spectrometer 
Cherenkov counter modified for the scattered 
particle envelope in the SOS.  This detector was filled with 
C$_4$F$_{10}$ and was pressurized 
to trigger on muons, but not on pions.  Additional muon 
identification was provided by requiring minimum 
ionizing pulse heights in each of the four layers of 
the lead glass shower counters. The overall efficiency for
muon pairs was estimated to be 80\%.

Drift chambers in each spectrometer were used to measure
particle momenta with a resolution of better than 0.2\%,
and scattering angles with an accuracy of 1 to 3 mr. The
resulting resolution on the di-lepton mass is approximately
10 MeV. Scintillator paddles in each spectrometer were
used for triggering, background rejection, and additional
particle identification. The last two features are illustrated
in Fig.~\ref{fig:coin}, which shows the number of events as a 
function of $\delta t$, the time difference between the
SOS and HMS relative to that expected for di-lepton
pairs (the time difference between electron-positron and
di-muon pairs is negligible on the scale used).  The top panel
(Fig.~\ref{fig:coin}a) is for the case of a cleanly identified
electron in the SOS, and all events in the HMS (dominated
by protons and pions). The peak near zero is due to pions,
and the peak near -4 ns is due to protons. The observed peak
widths are about 0.5 ns (1 $\sigma$).  The random
accidental background is very small. 

\subsection{Di-lepton events}
\begin{figure}[hbt]
\includegraphics[width=2.5in]{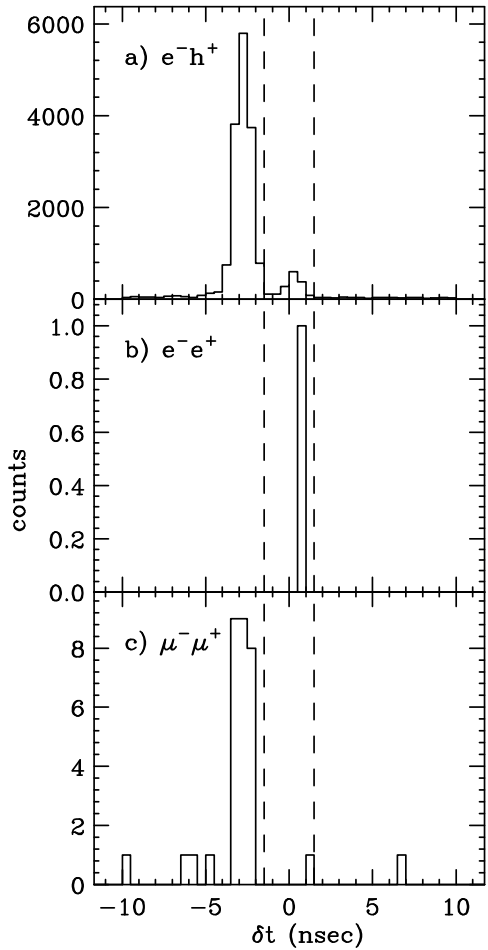}
\caption{Time difference spectra between the HMS and SOS
spectrometers, relative to that expected for di-lepton
events, for: a) electrons in the SOS and any particle in the HMS;
b) electrons in the SOS and positrons in the HMS; c) muons
in both the SOS and the HMS.}
\label{fig:coin}
\end{figure}

The electron-positron
time difference spectrum is plotted in Fig.~\ref{fig:coin}b. Only
one event is observed in the cut region $-1.5<\delta t<1.5$ ns
(illustrated as the vertical dashed lines), and there are no
events in the electron-proton or accidental coincidence regions.
The di-muon time difference spectrum 
is plotted in Fig.~\ref{fig:coin}c. Only
one event is observed in the cut region $-1.5<\delta t<1.5$ ns
but in this case there are 26 events in the lepton-proton
peak region. The latter are due to protons that produced a signal
in the Cherenkov counter (either from knock-on electrons, or from
scintillation light). A few accidental-in-time events are also visible. 

\begin{figure}[hbt]
\includegraphics[width=2.5in]{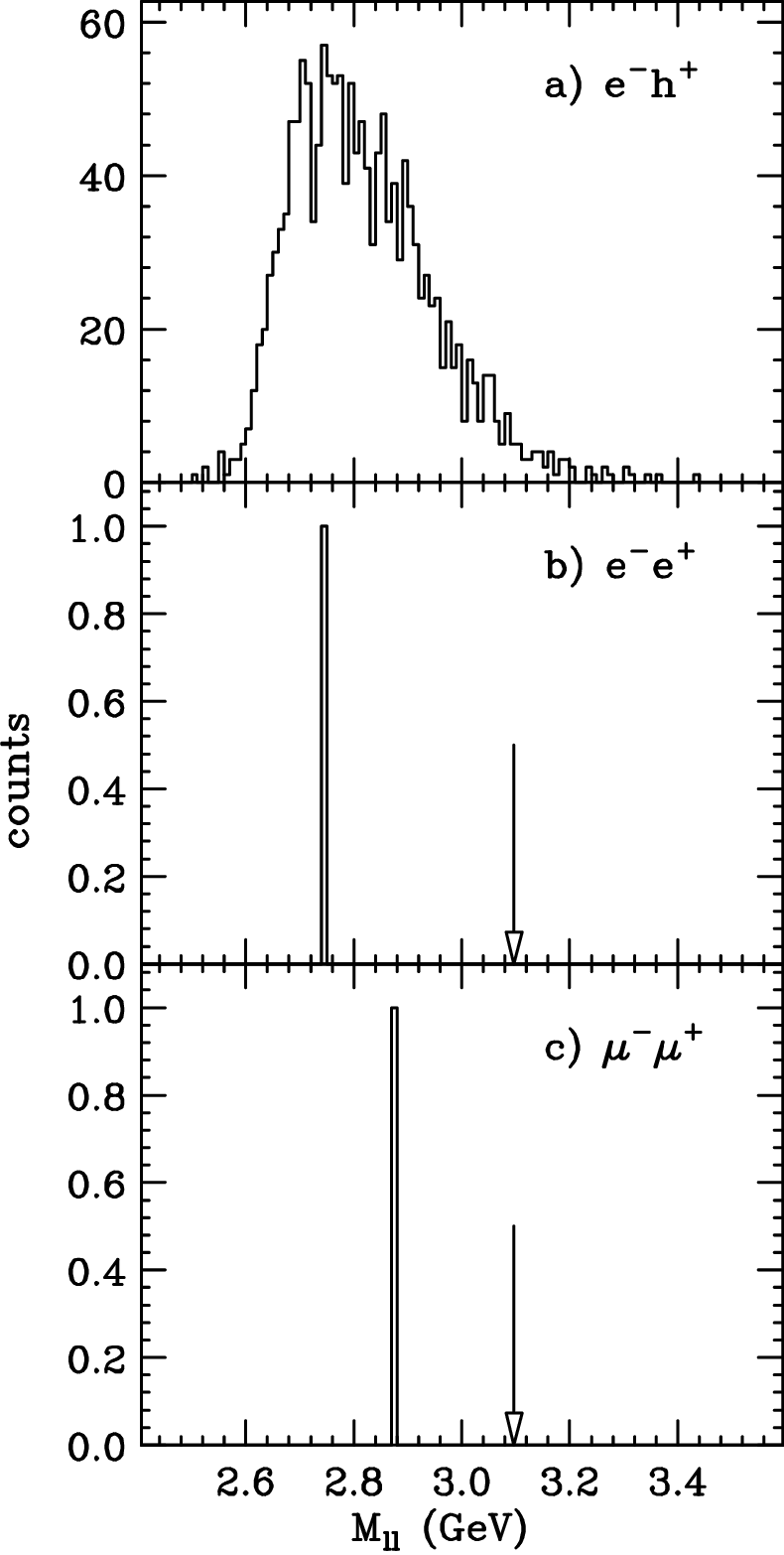}
\caption{Invariant mass spectrum from a negative
particle in the SOS, and a positive particle in the HMS, for
events passing the timing cuts shown in 
Fig.~\protect{\ref{fig:coin}}. The arrows correspond to the
\jpsi mass.}
\label{fig:mass}
\end{figure}

After putting a cut around the time-difference peaks shown
in Fig.~\ref{fig:coin}, the reconstructed
mass spectra using the HMS and SOS particles ($M_{ll}$)
was determined and are shown in Fig.~\ref{fig:mass}.  
 For electron-pion coincidences, shown in Fig.~\ref{fig:mass}a, 
the mass spectrum is smooth, and covers the mass range
$2.5<M_{ll}<3.5$ GeV. This basically illustrates 
 the phase space acceptance of the two spectrometers, and shows 
good acceptance at the \jpsi mass of 3.097 GeV.
As illustrated in Fig.~\ref{fig:mass}b, 
the one coincident $e^+/e^-$ event has an invariant 
mass of only $2.71\pm0.01$ GeV, more than 40 $\sigma$ from
the \jpsi mass. Thus this event is background: either an electron-pion
event (with the pion mis-identified as a positron), or a wide-angle
pair conversion of a Bremsstrahlung photon. 
As illustrated in Fig.~\ref{fig:mass}c, 
the one coincident $\mu^+/\mu^-$ event has an invariant 
mass of only $2.87\pm0.01$ GeV, more than 30 $\sigma$ from
the \jpsi mass. Thus this event is also background: 
mostly likely an accidental coincidence, but it could also be 
a wide-angle pair conversion of a Bremsstrahlung photon or a process
involving particle mis-identification, or the decay of pions to muons.

\subsection{\jpsi Acceptance}
In order to facilitate the calculation of the predicted number
of events for a given sub-threshold cross section model, we
give here the \jpsi acceptance corresponding to the HMS and SOS
setting used in this experiment. This was determined by a
detailed Monte Carlo simulation of the spectrometers, convoluted
with the di-lepton decays of \jpsi mesons with a particular 
momentum and polar angle with respect to the beam axis. 
The results are shown in Table~\ref{tab:acc}, and indicate
a maximum detection probability of about $0.5 \times 10^{-4}$ for
momenta near 4.3 GeV and angles less than 3 degrees.

\begin{table*}
\caption{\label{tab:acc} Calculated probability of detecting
a di-lepton pair from \jpsi decay in this experiment at the
indicated values of \jpsi momentum $P$  and lab angle
$\theta$ with respect to the beam direction. The results are 
scaled by $10^6$ and averaged over $e^+/e^-$ and $\mu^+/\mu^-$ pairs.}
\begin{ruledtabular}
\begin{tabular}{r|rrrrrrrrr}
$P$ (GeV) &
$\theta=0.4^\circ$ &
$1.2^\circ$ &
$2.0^\circ$ &
$2.8^\circ$ &
$3.6^\circ$ &
$4.4^\circ$ &
$5.2^\circ$ &
$6.0^\circ$ &
$6.8^\circ$ \\
5.81  &   0  &   0  &   2  &   4  &   0  &   2  &   4  &   0  &   0 \\     
3.95  &  78  &  58  &  84  &  76  &  44  &  30  &  24  &   8  &   5 \\
4.09  & 344  & 317  & 276  & 242  & 172  & 102  &  42  &  12  &   8 \\  
4.23  & 617  & 580  & 516  & 446  & 186  &  97  &  44  &  16  &  10 \\
4.37  & 478  & 440  & 394  & 337  & 243  &  91  &  48  &  12  &   6 \\
4.51  & 190  & 222  & 226  & 210  & 164  & 129  &  74  &  16  &   4 \\
4.65  &   8  &  26  &  51  &  63  &  82  &  48  &  26  &  16  &   2 \\
4.79  &   0  &   0  &   4  &   4  &   6  &   2  &   2  &   0  &   0 \\
\end{tabular}
\end{ruledtabular}
\label{accept}
\end{table*}

\subsection{Calibration Runs}
Several calibration runs were performed to check the 
kinematic settings and verify that the spectrometers were
fully functional.  The first calibration check was to set up 
the spectrometers for coincident $p(e,e^\prime,h)X$ 
running where $h$ could be either a positively-charged
pion, proton, or kaon.  A liquid hydrogen target 
replaced the carbon target 
and the HMS was moved forward to 13 degrees.  Three SOS 
angles were used, $53^\circ$, $40^\circ$, and $27^\circ$.  
These central kinematics corresponded to four momentum 
transfers of $Q^2 =$ 7.8, 4.6, and 2.1 GeV$^2$, 
and invariant mass $W = 0.8$, 2.0, and 2.5 GeV, respectively.  
In all cases, the peaks in the missing mass spectra were at 
the proper positions and the widths were understood.  As 
an example, using proton cuts on the HMS arm, the missing 
mass spectrum for $p(e,e^\prime,p)X$ shows prominent $\eta$ 
and $\pi^0$ peaks at the correct locations (as seen in previous
Hall C experiments~\cite{Dalton}).  As a further check of 
the luminosity, accidental and multi-pion continuum backgrounds 
were subtracted from the data to produce experimental yields.  
These yields were corrected for radiative effects and 
spectrometer acceptances to produce differential cross sections.  
Good agreement was found between the measured and simulated 
differential cross sections~\cite{Dalton}.  
This indicates that the target 
density and electron beam current determinations were well understood.

\begin{figure}[hbt]
\includegraphics[width=2.5in]{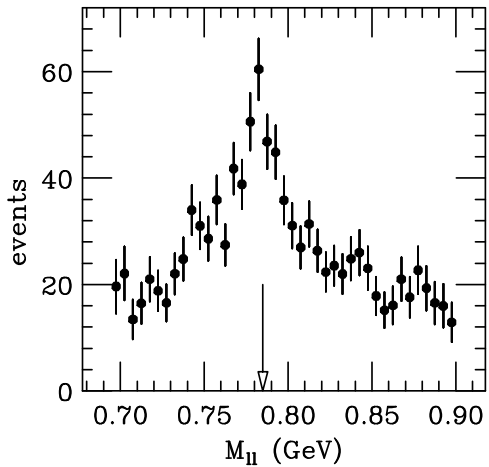}
\caption{Acceptance-corrected invariant mass spectrum from an electron
in the SOS and a positron in the HMS, at kinematics optimized 
for detection of $\omega(785)\rightarrow e^+e^-$ (indicated by
arrow) and $\rho(770)\rightarrow e^+e^-$.}
\label{fig:om}
\end{figure}

The other significant calibration check was to set the 
spectrometers to measure lepton pair decays from other 
lower mass vector mesons.  In particular, we centered the 
spectrometers to detect  $e^-/e^+$ pairs from  $\omega$ and $\rho$
meson decays.  For these runs a beryllium target was used with the 
SOS set to 25 degrees and a central momentum of 0.9 GeV, 
and the HMS set to 11 degrees and a central momentum of 
1.95 GeV.  In spite of the very small branching ratios of
about $0.5\times10^{-5}$, a clear $\omega(785)$ peak with the expected
full width of 8 MeV was seen on top of a 140 MeV wide $\rho(770)$ peak,
as illustrated in Fig.~\ref{fig:om}.
The number of detected $\omega$ events is consistent with the 
predicted number from a Monte Carlo simulation that had as much
as possible in common with the simulation of \jpsi events.

\section{Results}

\begin{figure*}[hbt]
\begin{center}
\includegraphics[width=7in]{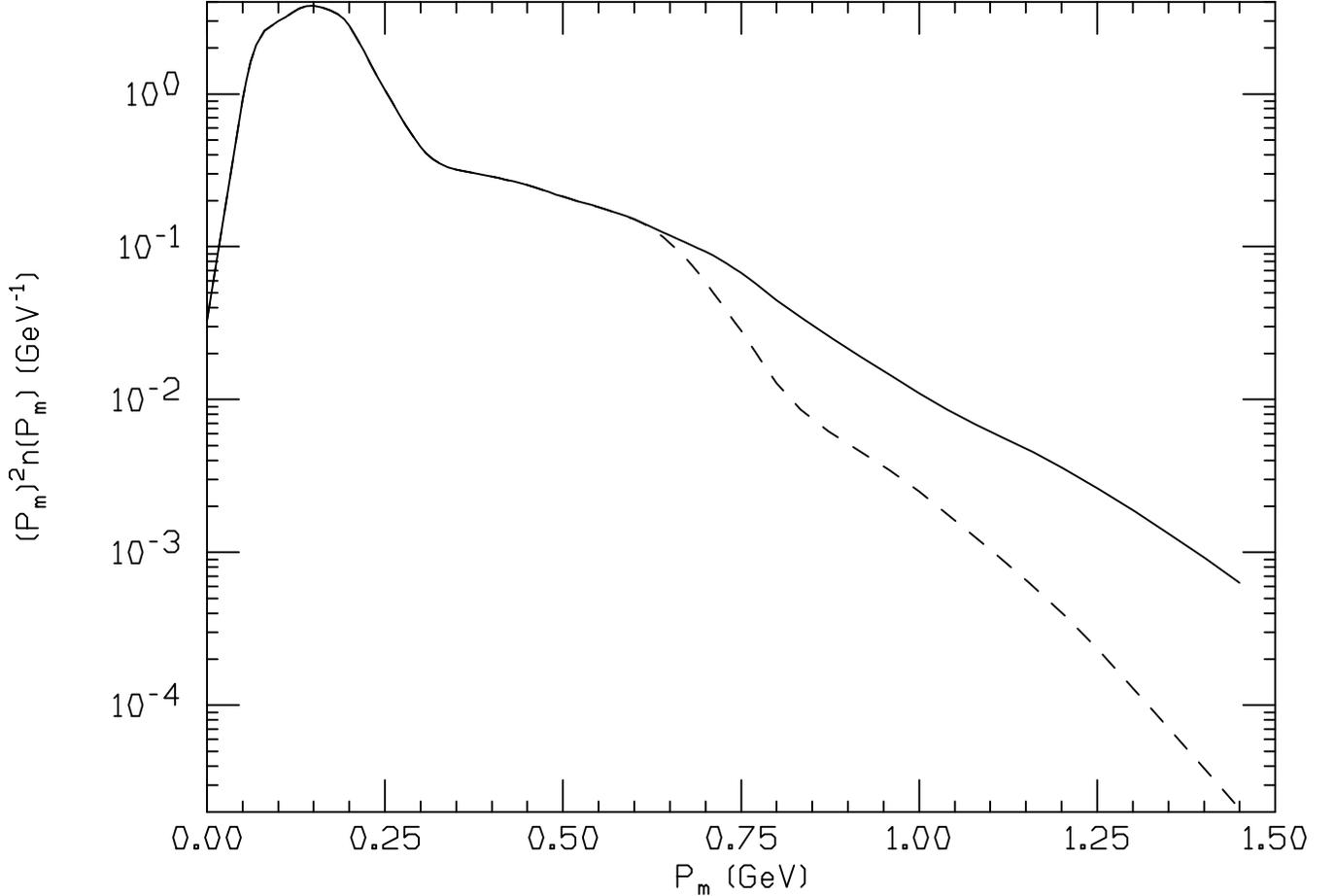}
\caption{The $E_m$-integrated probability of finding a nucleon
with missing momentum $P_m$ as a function of $P_m$. The solid
(dashed) lines are ``high'' and ``low'' eyeball extrapolations 
above 0.6 GeV of the spectral function 
of Ref.~\protect{\cite{Benhar}}. }
\label{fig:sf}
\end{center}
\end{figure*}

Since no \jpsi mesons were detected in this experiment, 
here we try to quantify what the implications are for
elementary  photoproduction cross section models, in the
context of the single-particle convolution model described
above, for four choices of the prescription used to extend the
spectral function of Benhar~\cite{Benhar} beyond $P_m=0.6$ GeV (where
there is insufficient data from $C(e,e^\prime p)X$ to constrain the
function).
We show two ``eyeball'' extrapolations of the probability
distribution integrated over $E_m$ in Fig.~\ref{fig:sf}. We
label this as ``high'' and ``low''  according whether
they are are higher or lower at large $P_m$. As it happens, these
extrapolations are quite similar to the two extrapolations (for
nuclear matter) shown in Fig.~1 of Ref.~\cite{Hipm}. As discussed
in Ref.~\cite{Hipm}, the ``high'' extrapolation should probably
be considered as an upper limit, and probably will over-estimate
yield predictions. We
used two choices for the $E_m$ distribution: a) ``freeze'' the
$E_m$ distribution to that of Benhar et al.~\cite{Benhar} 
at $P_m=0.8$ GeV (the highest
provided in this fit); b) shift the $E_m$ distribution by
$\sqrt{m^2 + P_m^2} - \sqrt{m^2 + 0.8^2}$, with $P_m$ in units of
GeV. The latter corresponds to following the ``ridge'' observed at
lower $E_m$. Using the nuclear matter prescription of 
Eq.~10 in Ref.~\cite{Hipm} should give results somewhere between
these two cases, as their prescription involves shifting the
$E_m$ distribution along the same ``ridge'', but having the
width increase with increasing $P_m$. 
Following Ref.~\cite{Strikman}, we also investigated 
scaling the spectral function by a
relativistic correction factor $m/(m-E_m)$ and found the results
to change by less than 20\%.

\begin{figure*}[hbt]
\begin{center}
\includegraphics[width=7in]{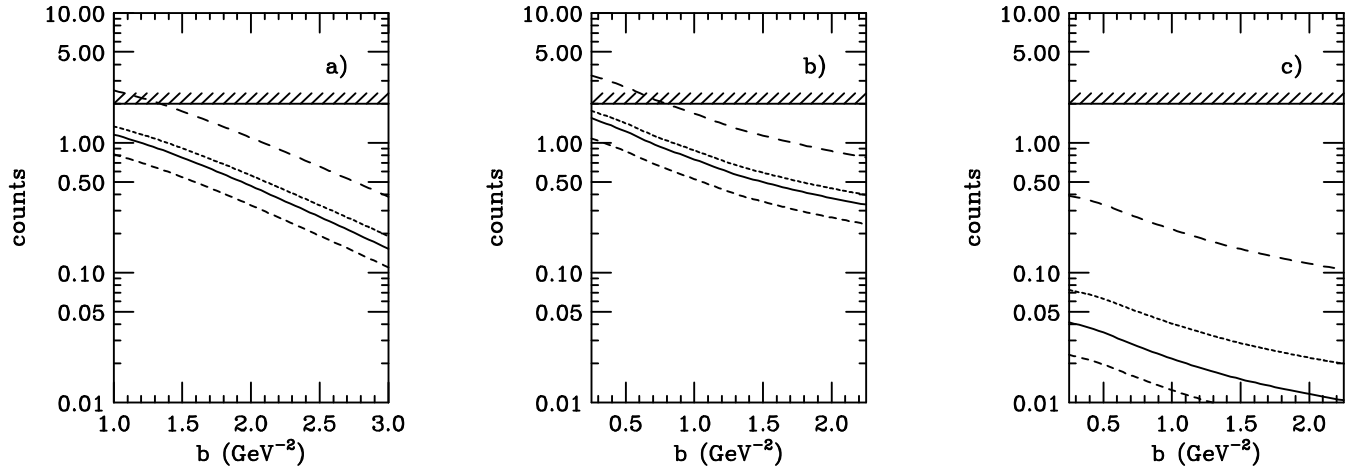}
\caption{The number of predicted counts for the conditions of the
present experiment as a function of $b$ for: a) Model I; b) Model II;
and c) Model III, where the models are defined in 
Equations 4, 5, and 6, respectively.
On each panel, the solid (long dashed) curves use the ``high''
extrapolation of the spectral function (see Fig.~\ref{fig:sf})
with (without) the $E_m$ shift for $P_m>0.8$ GeV described in the
text. The medium dashed (short dashed) are the corresponding curves
with the ``low'' extrapolation. The hatched band at 2 counts 
indicates the 87\% confidence level corresponding to our 
experimental observation of no events.}
\label{fig:limits}
\end{center}
\end{figure*}

We used three different free nucleon cross sections (as motivated by
the discussion in the introduction):
\begin{eqnarray}
\label{eq:Models}
\mbox{I.}\ \ d\sigma/dt &=& a e^{bt} \\
\mbox{II.}\ \  d\sigma/dt &=& a / (1 - bt)^4  \\
\mbox{III.}\ \  d\sigma/dt &=& a (1-x)^2 /(1 - bt)^4,  
\end{eqnarray}
 where $a$ and $b$ are free parameters, and 
we used $x=((m+M_J)^2-m^2)/(s-m^2)$ 
For each model, we varied the $t$-slope parameter $b$ within a
reasonable range and, for each value of $b$, determined $a$ such that the
total cross section would agree with the Cornell 
measurement~\cite{Cornell} of 0.7 nb at $k=11$ GeV.
The predicted counts are shown in Fig.~\ref{fig:limits} as
a function of $b$ for each of the three models for four high-$P_m$
extrapolations of the spectral function. The predictions for 
model III are quite a bit lower than models I and II, due to the
factor of $(1-x)^2$. The sensitivity to the spectral function
extrapolation is also largest for model III, because on average
higher values of $P_m$ are probed. For all models, ``shifting'' the
$E_m$ distributions for $P_m>0.8$ GeV makes a difference of 
typically a factor of two in predicted rates. Not shown in the
figure is the effect of assuming $E_m=P_m^2/2(A-1)m$ (i.e., a virtual photon
interacting with an almost on-shell nucleon), as was initially assumed
in the planning phase of the present experiment. This assumption
results in predicted rates higher by approximately two orders of
magnitude than if a more realistic $E_m$ distribution is used. 

While the predicted number of counts is below our 87\%
confidence level observation of less than 2 counts (shown as the
hatched bands in Fig.~\ref{fig:limits}) for all the combinations
considered here, it is not impossible that
for some cross sections models (I and II in particular), something
like the ``hot spot'' scenario discussed in the introduction could
result in a prediction of more than two counts. In the quasi-free
picture, this could correspond to considerably lower average
values of $E_m$ than in the Benhar spectral function.

While we are not aware of any detailed theoretical 
predictions for sub-threshold \jpsi photoproduction specifically, we can 
use the calculation of charm ($C\bar C$) photoproduction of 
Braun and Vlahovic~\cite{Braun} as a possible guide. Using pQCD to evaluate
the photon-gluon fusion process, they predict a total cross section
of 0.25 fb/nucleon for carbon for 5.5 GeV photons. This 
corresponds to approximately 0.02 events for our experimental conditions. 
This is similar to our predictions using Model III.

\section{Conclusion}

The non-observation of sub-threshold \jpsi photoproduction on
carbon in the present experiment is consistent with predictions
of quasi-free production with a variety of reasonable 
elementary free nucleon cross
sections models near threshold, and educated guesses for the
high missing momentum and missing energy spectral function
distributions in carbon. For a given set of assumptions on 
 the cross section and spectral
function choices, upper limits could be set on
exotic mechanisms that could potentially 
enhance the sub-threshold cross sections (such as gluon exchange to
two different nucleons, hidden color configurations, 3-gluon
exchange, etc.).  The interpretation of the present experiment will
be greatly aided with precision measurement of the elementary 
free nucleon cross section near threshold,  
planned at Jefferson Lab~\cite{E1207} once beam 
energies up to 11 GeV are available. 

The authors wish to thank J.M.~Laget 
for theoretical support. This work is supported in part by research 
grants from the U.S. Department 
of Energy and the U.S. National Science Foundation. The South African
group was acknowledges the support of the National Research Foundation.
The Southeastern Universities Research Association operates the
Thomas Jefferson National Accelerator Facility under the
U.S. Department of Energy contract DEAC05-84ER40150.

\end{document}